\documentclass[useAMS,usenatbib,twocolumn]{mn2e}
\usepackage{epsfig,graphicx}
\usepackage{amssymb}
\usepackage{amsbsy}

\title{Microwave sky and the local Rees--Sciama effect}
\author[Aleksandar Raki\'c, Syksy R\"as\"anen and Dominik J. Schwarz]
{Aleksandar Raki\'c$^{1}$\thanks{E-mail: rakic at physik dot uni-bielefeld dot de},
 Syksy R\"as\"anen$^{2}$\thanks{E-mail: syksy dot rasanen at iki dot fi}
 and
 Dominik J. Schwarz$^{1}$\thanks{E-mail: dschwarz at physik dot uni-bielefeld dot de}\\
 $^{1}$Fakult\"at f\"ur Physik, Universit\"at Bielefeld, Postfach 100131,
  D-33501 Bielefeld, Germany\\
 $^{2}$CERN, Physics Theory Department, CH-1211 Geneva 23, Switzerland}

\begin{document}

\date{Accepted 2006 March 9. Received 2006 March 1; in original form 2006 January 25}

\pagerange{\pageref{firstpage}--\pageref{lastpage}} \pubyear{2006}

\maketitle

\label{firstpage}

\begin{abstract}
The microwave sky shows unexpected features at the largest angular scales, 
among them the alignments of the dipole, quadrupole and octopole. 
Motivated by recent X-ray cluster studies, we investigate the possibility 
that local structures at the 100~$h^{-1}$Mpc scale could be responsible 
for such correlations. These structures give rise to a local Rees--Sciama 
contribution to the microwave sky that may amount to $\Delta T/T \sim 
10^{-5}$ at the largest angular scales. We model local structures by a 
spherical overdensity (Lema\^itre--Tolman--Bondi model) and assume that 
the Local Group is falling toward the centre. We superimpose the local 
Rees--Sciama effect on a statistically isotropic, gaussian sky. As expected,  
we find alignments among low multipoles, but a closer look reveals that 
they do not agree with the type of correlations revealed by the data.
\end{abstract}


The microwave sky has presented some surprises at the largest angular scales.
The Wilkinson Microwave Anisotropy Probe (WMAP) confirmed the 
vanishing of the angular two-point correlation function above $60^\circ$ 
\citep{Bennett:2003a}, a result first obtained by the Cosmic Background
Explorer's Differential Microwave Radiometer (COBE-DMR) experiment
\citep{Hinshaw:1996}. In terms of the angular power spectrum this
implies that the quadrupole and octopole are below the 
theoretical expectation.

\hyphenation{fore-gr-ound-clean-ed}
\hyphenation{fu-ll-sky}
The analysis of full-sky maps that have been cleaned from foreground
\citep*{Bennett:2003b},\citep*{Tegmark:2003},\newline \citep*{Eriksen:2004b} has revealed further
surprises. It was pointed out by \citet{deOliveira-Costa:2004} that the octopole
seems to be planar (all minima and maxima are close to a great
circle on the sky) and the planes of the octopole and the quadrupole are
closely aligned. \citet{Eriksen:2004a} showed that the northern galactic
hemisphere lacks power compared with the southern hemisphere.
By means of multipole vectors \citep*{Copi:2004}, 
\citet{Schwarz:2004} showed that the quadrupole and octopole
are correlated with each other and
with the orientation and motion of the Solar system. 
The four cross products of the quadrupole and octopole vectors 
are unexpectedly close to the ecliptic [$> 98\%$ confidence level (C.L.)] 
as well as to the equinox (EQX) and microwave dipole (both $> 99.7\%$ C.L.)
\citep{Copi:2005}. Based on the additional alignment of a nodal line with
the ecliptic and an ecliptic north--south asymmetry of the quadrupole plus 
octopole map, \citet{Copi:2005} argued that the correlation with the
ecliptic is unlikely at the  $> 99.9\%$ C.L. In contrast to an unknown Solar
system effect, it also seems possible
that the large-scale anomalies are due to a physical correlation with the
dipole, in which case the correlation with the ecliptic and the EQX 
would be due to the accidental closeness of the dipole and the EQX.

In this letter we explore the possibility that the effect of local 
non-linear structures on the cosmic microwave background (CMB), 
the local Rees--Sciama (RS) effect \citep{Rees:1968}, could induce a 
correlation between the dipole and higher multipoles. In the non-linear 
regime of large-scale structure formation
the gravitational potential changes with time, and photons climb 
out of a potential well slightly different from the one they fell into.
As the CMB dipole is considered to be due to our motion
with respect to the CMB rest frame, and this motion is
due to the gravitational pull of local structures, these structures
are a natural candidate for contributions to the higher multipoles
correlated with the dipole. 
Earlier work on the effects of large nearby structures on the CMB includes the
moving cluster of galaxies effect \citep{Vale:2005, Cooray:2005}, which
is different from the effect we are considering here. Second order corrections to
the integrated Sachs-Wolfe effect, which catch some aspects of the local RS
effect have been considered by Tomita (2005a, 2005b).

The RS effect of distant clusters was estimated
to be at most $10^{-6}$ in a matter-dominated Universe
in \citet{Seljak:1996}, one order
of magnitude below the intrinsic CMB anisotropy. The effect of local
large structures has been estimated to be at most $10^{-6}$
using the Swiss Cheese model \citep{MG:1990} and, more reliably,
the Lema\^itre--Tolman--Bondi (LTB) model,
which is the general spherically symmetric dust solution
of the Einstein equation \citep*{Panek:1992, Saez:1993, Fullana:1994}.
For an overview, see \citet{Krasinski:1997}.

\hyphenation{Rad-bu-rn-Smith}
At the time these studies
were made, it was generally thought that the dipole is mostly
due to the infall of the Local Group (LG) of galaxies towards the Great
Attractor (GA) \citep{LB:1988, Dressler:1988}, a density
concentration located 40--60~$h^{-1}$Mpc from us,
with a subdominant component due to the nearby Virgo cluster, 
about 10~$h^{-1}$Mpc away.
Recent observations of X-ray clusters suggest instead that
there is a major contribution to the dipole from the Shapley
Supercluster (SSC) and other density concentrations 
at a distance of around 130--180~$h^{-1}$ Mpc 
\citep*{Kocevski:2004, Hudson:2004, Lucey:2004, Kocevski:2005}.
The SSC alone has a density contrast of $\approx$ 5 over a
30~$h^{-1}$Mpc region \citep{Proust:2005}, which is 2--3 times the size of the
core (of similar density) in the GA models. 

\hyphenation{velo-city}
An SSC-like object could induce anisotropies at the $10^{-5}$ level,
consistent with an early estimate in \citet{MG:1990}. This can be understood by
the approximate scaling \citep{Panek:1992} 
\begin{equation} 
\label{RS}
 \Delta T/T \sim \delta^{3/2} (d/t)^3 \, ,
\end{equation}  
where $\delta$ is the density contrast of the structure, $d$ is its
size and $t$ is the time at which the CMB photons crossed it.
For a large angular scale of the source (local and nearby structures), this
induces contributions to low-$\ell$ multipoles, especially the dipole, 
quadrupole and octopole.  
This could include a non-Doppler contribution to the dipole. This would imply a
change of a few percent in the inferred dipole velocity, which might also
explain some of the CMB anomalies \citep{Freeman:2005}. 
The SSC is a non-linear structure, and the amplitude of the
induced anisotropies cannot be reliably calculated in linear
perturbation theory. According to a comparison of linear and
exact calculations for GA-like objects with the LTB model
in \citet{Fullana:1994}, linear theory is reliable
at distances comparable to the Hubble scale, but fails
for structures within 1000~$h^{-1}$Mpc or so.

The advantage of the spherical symmetry of the LTB model is that
it allows exact calculations for non-linear objects;
the drawback is that the observed non-linear objects
such as the GA and SSC do not appear to be spherically symmetric.
However, one would expect the result to be correct within
an order of magnitude, and the core of the SSC 
does seem to be roughly spherical \citep{Proust:2005}.
Also, if the preferred direction indicated by the low-$\ell$ 
anomalies is due to local structures, this implies that
there indeed \emph{is} a degree of symmetry in the local mass distribution.

In addition, there is a second motivation for studying a
spherically symmetric inhomogeneous model, namely dark energy.
If interpreted in the framework of isotropic and homogeneous cosmology,
observations of SNIa imply that the expansion of the Universe is
accelerating. However, in an inhomogeneous 
spacetime the observations are not necessarily inconsistent with
deceleration. In particular, in the LTB model the parameter
$q_0$ defined with the luminosity distance is no longer
a direct measure of acceleration \citep*{Humphreys:1997}.
It has been suggested by several groups that a spherically
symmetric inhomogeneity could be used to explain the SNIa data
\citep*{Celerier:1999, Tomita:2001, Alnes:2005}, though
it is not clear whether such a model could be consistent with what is
known about structures in the local Universe \citep{Bolejko:2005}
or the observation of baryon oscillations in the matter power spectrum.
Here we concern ourselves only with the CMB.

The picture of the local Universe that we adopt is a spherically 
symmetric density distribution, with the LG falling towards the core of 
the overdensity at the centre. The line between our location
and the centre defines a preferred direction $\hat{\bf z}$, which in the
present case corresponds to the direction of the dipole
after subtracting our motion with respect to the LG and the LG's infall 
towards the nearby Virgo cluster (assuming the primordial component of 
the dipole to be negligible). The directions on the sky that are important 
for our analysis are given in Table \ref{table1}. This setup exhibits
rotational symmetry w.r.t.~the axis $\hat{\bf z}$ (neglecting transverse
components of our motion). Consequently, only zonal
harmonics ($m=0$ in the $\hat{\bf z}$-frame) are generated.

\begin{table}
\centering
\caption{Directions of local motion with respect to the CMB rest frame. 
  Estimated error for the corrected LG direction of 
  \citet{Plionis:1998}(PK) is $14^\circ$, and is 5\% for the velocities.}   
\begin{tabular}{llc}
\hline 
  Direction & Galactic coordinates & $V$ [km/s] \\
\hline
\hline
 WMAP dipole velocity & $l=263^\circ\!\!.85 \pm 0^\circ\!\!.10$ & (368$\pm$2)\\ 
 $[$\citet{Bennett:2003a}$]$ & $b=48^\circ\!\!.25 \pm 0^\circ\!\!.04$ & \\[1mm]
 LG velocity & $l=276^\circ \pm 3^\circ$ & (627$\pm$22) \\
 $[$\citet{Kogut:1993}$]$ & $b=30^\circ \pm 3^\circ$ & \\[1mm]
 Virgo infall of LG & $l=283^\circ\!\!.92$ & 170 \\
 $[$PK (1998)$]$ & $b=74^\circ\!\!.51$ & \\[1mm]
 Virgo corrected LG vel. & $l=276^\circ$ & 510 \\ 
 $[$PK (1998)$]$ &  $b=16^\circ$ & \\[1mm]
 Shapley concentration & $l=306^\circ\!\!.44$ & - \\
 $[$\citet{Einasto:1997}$]$ & $b=29^\circ\!\!.71$ & \\[1mm]
\hline
\label{table1}
\end{tabular}
\end{table}  

The density field has two effects on the CMB seen by
an off-centre observer. First, photons coming from different
directions travel different routes through the local
overdensity, and this creates anisotropy (even with a perfectly
homogeneous distribution of photons). In a stationary setup
(i.e. for virialised structures) this effect vanishes and there
is no imprint on the CMB. Second, the environment will
affect the evolution of the intrinsic anisotropies (as
the homogeneous background space does, by changing the
angular diameter distance). The correct calculation that
would account for both of these effects would be to study
the evolution of the CMB anisotropies as they travel across
the density field using perturbation theory on the LTB background.
We will present the calculation of the amplitude of the anisotropies
induced by a local structure described with the LTB model elsewhere.  
As in earlier treatments, we neglect the second effect
and simply add the anisotropy
generated by the LTB model on top of the intrinsic
contribution. It is possible that this treatment misses
some effects of processing the anisotropies already present.
In particular, simply linearly adding a new source of anisotropy will in
general add multipole power, not reduce it, while a proper analysis
of the processing of the intrinsic anisotropies could lead to a
multiplicative modification of the amplitudes of the low multipoles,
discussed in \citet{Gordon:2005}.
\begin{figure*}
\centerline{\hfill 
\includegraphics[height=3.1in, angle=270]{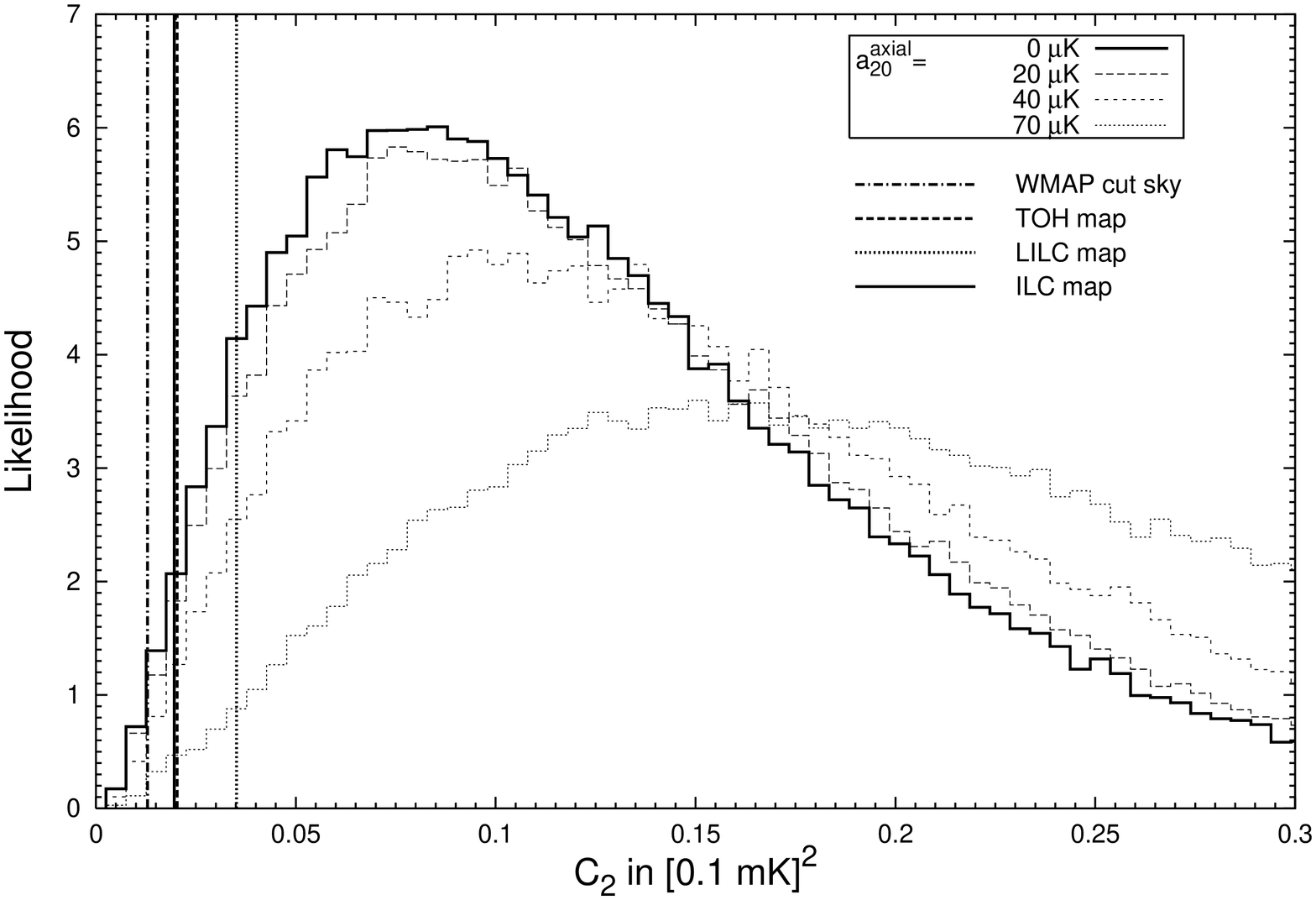}\hfill
\includegraphics[height=3.1in, angle=270]{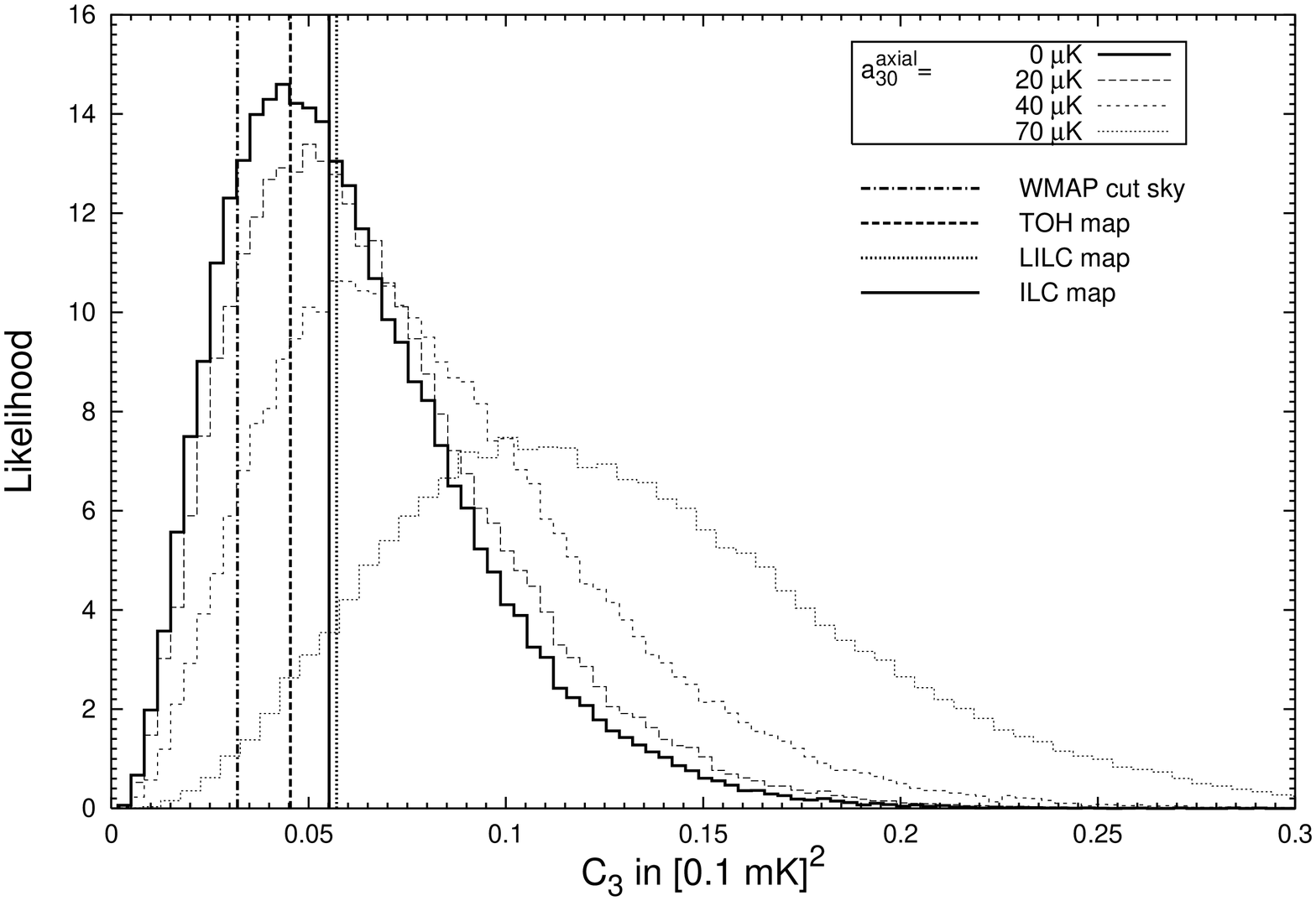}\hfill}
\caption{Likelihood of quadrupole and octopole power with increasing axial 
  contamination.  
  Vertical lines indicate measured values as given in Table \ref{table2}. 
  From the WMAP cut-sky analysis, adding \emph{any}
  multipole power to the quadrupole is already excluded at $> 99\%$C.L., 
  whereas it is possible to add up to $80\ \mu$K to the octopole until 
  reaching the same exclusion level.} \label{fig1}
\end{figure*}

Any other effect with axial symmetry would also induce anisotropy only 
for the $m=0$ components.
It has been suggested that spherically symmetric inhomogeneities of the
order of horizon size or larger would contribute to the low CMB multipoles 
\citep{Grishuck:1978, Raine:1981, Paczynski:1990, Langlois:1996};
it was claimed in \citet{Moffat:2005} that this could explain   
the observed preferred axis. Leaving aside the issue that
assuming spherical symmetry for the entire Universe seems questionable,
the observational signature on the low multipoles is identical to that
from the LTB model used to describe local structures, possibly apart from 
the amplitude.

We study how the CMB is affected by the anisotropy induced
by the additional axisymmetric contribution by means of Monte Carlo 
(MC) maps. We keep the amplitudes as free parameters to cover both
local and horizon-sized structures at the same time,
and look at how the observational signature compares with what is
actually seen on the sky.

The angular power spectrum is given by $C_\ell = \sum_m |a_{\ell m}|^2/(2\ell+1)$, 
where $a_{\ell m}$ are the coefficients in the
harmonic decomposition of the temperature map.
As predicted by the simplest inflationary models, we assume that the 
$a_{\ell m}$ are statistically independent (isotropy), 
gaussian and have zero mean. The $a_{\ell m}$ are then fully 
characterised by angular power.
We use the values $C_2=1278.9\ \mu$K$^2$, 
$C_3=590.9\ \mu$K$^2$ from the best-fitting temperature spectrum
of the $\Lambda$ cold dark matter ($\Lambda$CDM) model with a power-law
primordial perturbation spectrum 
to the combined WMAP, Cosmic Background Imager (CBI) and Arcminute Cosmology
Bolometer Array Receiver (ACBAR) dataset.\footnote{WMAP data products 
at http://lambda.gsfc.nasa.gov/} 
For the statistical analysis we generate $10^5$ MC realizations 
of the quadrupole and the octopole. 
As the contribution of the structure described by the
LTB model, we add to the quadrupole and the octopole
a component, denoted by $a_{\ell 0}^{\rm{axial}}$, which is a pure
$m=0$ mode with respect to a given physical direction 
$\hat{\bf z}$.

First we address the amplitude of the quadrupole and
the octopole. The values of $C_2$ and $C_3$ determined from the
WMAP cut-sky \citep{Hinshaw:2003}, 
the TOH map \citep{Tegmark:2003}, 
the Lagrange ILC map \citep{Eriksen:2004b} and the Internal Linear Combination
(ILC) map \citep{Bennett:2003b} are listed in Table \ref{table2}. The extracted 
quadrupoles have been Doppler-corrected 
as described in \citet{Schwarz:2004}, except for the cut-sky
value. The values of $C_2$ and $C_3$ from the full-sky maps are
significantly larger than the cut-sky values.

Figure \ref{fig1} shows how the $C_2$ and $C_3$ histograms compare with the data
as $a_{\ell 0}^{\rm{axial}}$ is increased. For $a_{\ell 0}^{\rm{axial}}=40\ \mu$K, the
number of MC hits that are consistent with the WMAP cut-sky data is smaller by a
factor of $\sim 2$ for both $C_2$ and $C_3$ compared with the pure CMB sky. For
$a_{\ell 0}^{\rm{axial}} = 70\ \mu$K, the number of consistent MC hits for $C_2
(C_3)$ is reduced by a factor of $\sim 5 (15)$ compared with the pure CMB sky. 
Note that adding \emph{any} power to the theoretically expected quadrupole 
is excluded at the $> 99\%$C.L. level from the cut-sky analysis, but
for the octopole the same exclusion level is not reached
until $a_{30}^{\rm{axial}}=80\ \mu$K. 

Next we ask what kind of directional patterns the contribution
$a_{\ell 0}^{\rm{axial}}$ induces on the sky.
In the multipole vector representation \citep{Copi:2004} any
real multipole $T_\ell$ on a sphere (with radial unit vector $\hat{\bf e}$) can be
expressed with $\ell$ unit vectors $\hat{{\bf v}}^{(\ell,i)}$ and one scalar $A^{(\ell)}$ as 
\begin{equation}
  T_\ell = \sum_{m=-\ell}^\ell a_{\ell m} Y_{\ell m}(\theta, \phi) \simeq 
  A^{(\ell)} \prod_{i=1}^\ell \hat{\bf v}^{(\ell,i)} \cdot \hat{\bf e} \, .
\label{eq_multipoles}
\end{equation}
\noindent
The signs of the multipole vectors can be absorbed into the scalar
quantity $A^{(\ell)}$, and are thus unphysical.
Note that in (\ref{eq_multipoles}) the r.h.s.~contains contributions
with angular momentum $\ell-2$, $\ell-4$, \dots 
The uniqueness of the multipole vectors is ensured by
removing these terms by taking the
appropriate traceless symmetric combination \citep{Copi:2004}.
Note that the multipole vectors are independent of the angular power. 
As in \citet{Schwarz:2004}, we introduce 
the $\ell(\ell-1)/2$ oriented areas ${\bf w}^{(\ell;i,j)} \equiv 
\hat{{\bf v}}^{(\ell,i)} \times \hat{{\bf v}}^{(\ell,j)}$.
Alignment of the normals
${\bf n}^{(\ell;i,j)} \equiv \pm \, \hat{\bf w}^{(\ell;i,j)}$
with a given direction $\hat{\bf x}$ is tested with
\begin{equation} 
 S_{\rm{{\bf nx}}} \; \equiv \; \frac{1}{4} \, \sum_{\ell=2,3} 
 \sum_{i < j} \left| {\bf n}^{(\ell;i,j)} \cdot \hat{{\bf x}} \right| \, . 
 \label{eq_Snx}   
\end{equation}  
This statistic is a sum over all dot products for a given
$\hat{{\bf x}}$, so it does not imply any ordering between the terms and is
a unique and compact quantity. For computing the multipole vectors we
use the method introduced by
\citet{Copi:2004}.\footnote{Code at http://www.phys.cwru.edu/projects/mpvectors/}

\begin{figure*}
  \includegraphics[width=1.64in,angle=-90]{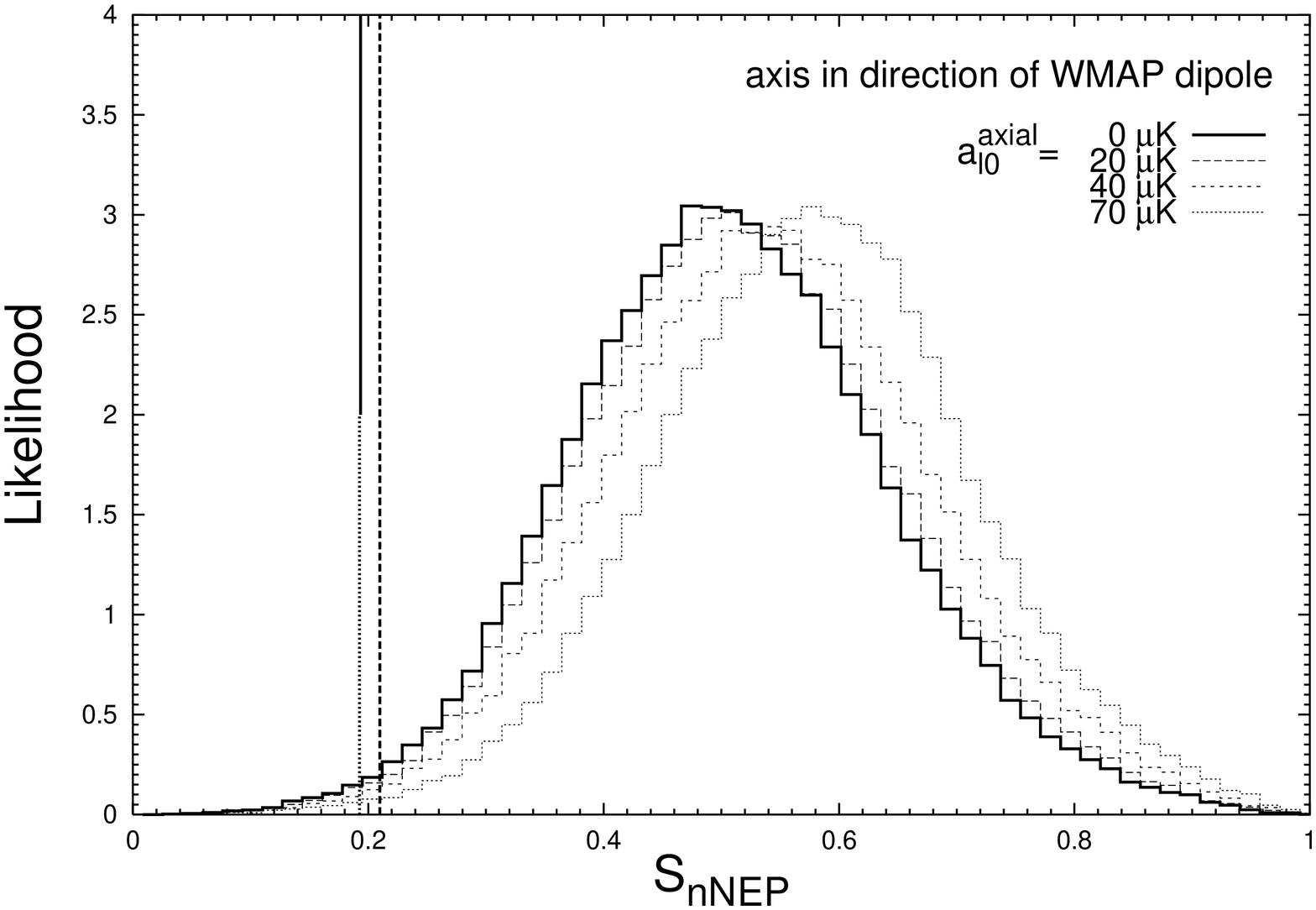}\hspace{-0.2cm}
  \includegraphics[width=1.64in,angle=-90]{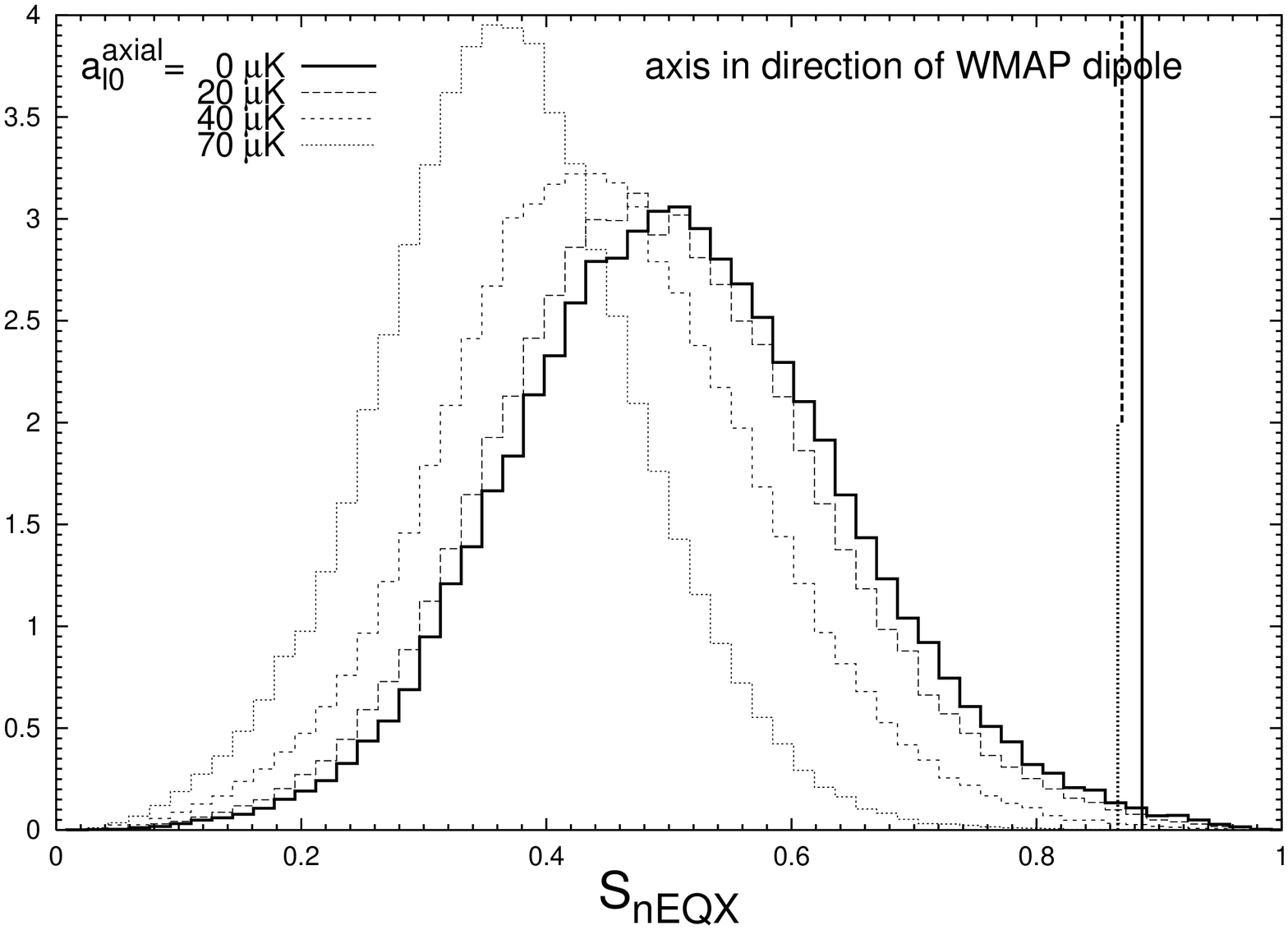}\hspace{-0.2cm}
  \includegraphics[width=1.64in,angle=-90]{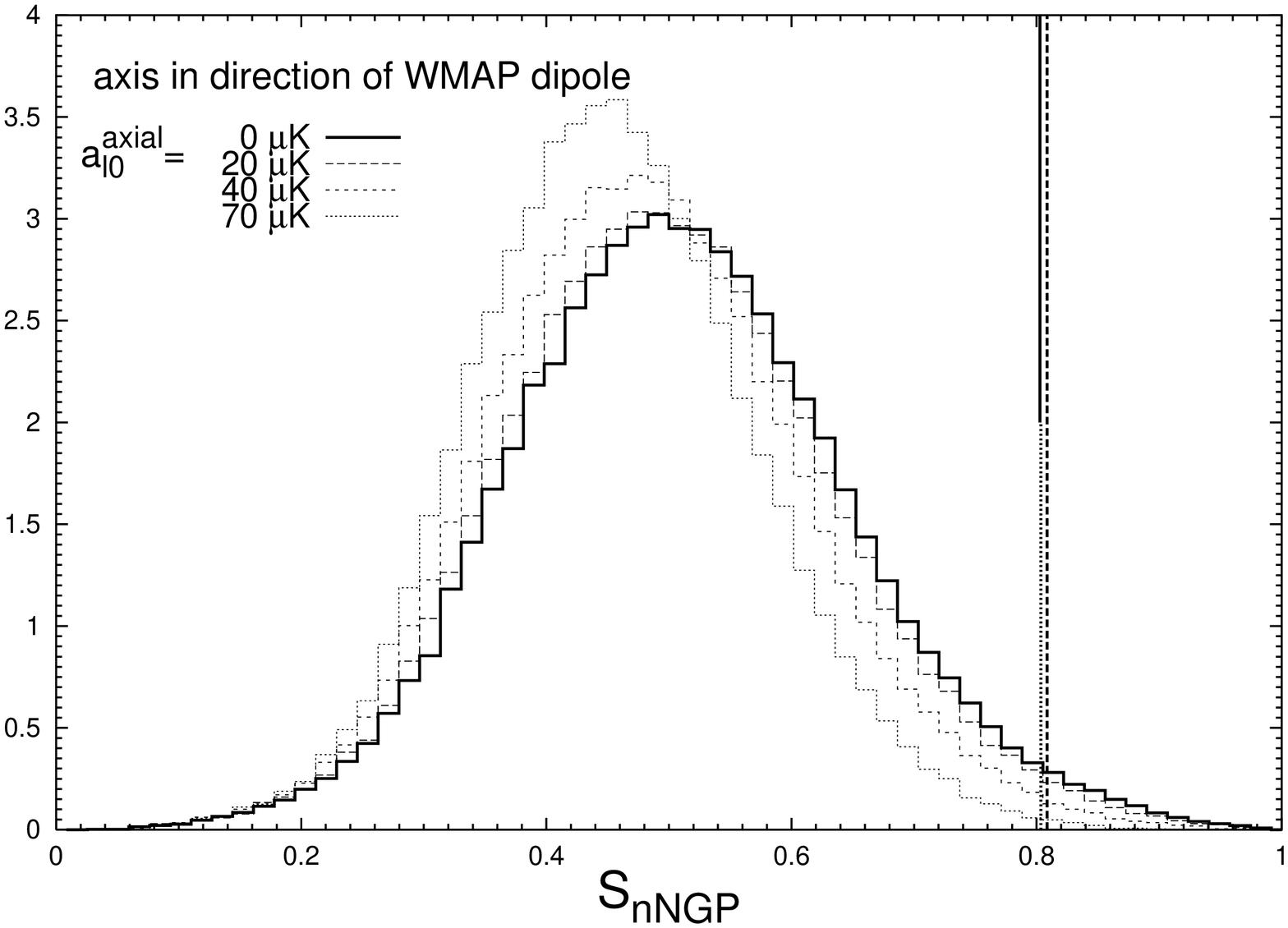}\\
  \includegraphics[width=1.64in,angle=-90]{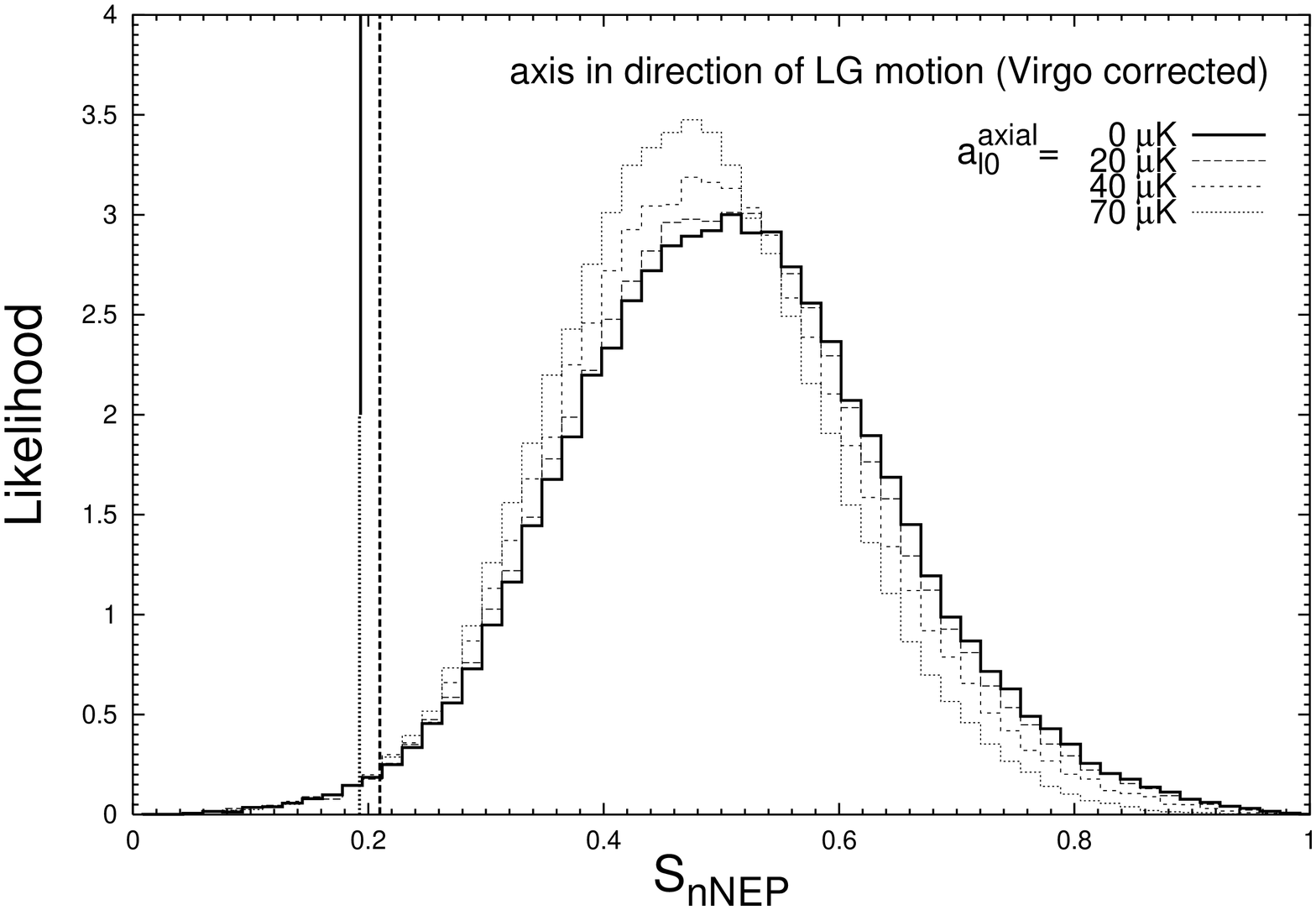}\hspace{-0.2cm}
  \includegraphics[width=1.64in,angle=-90]{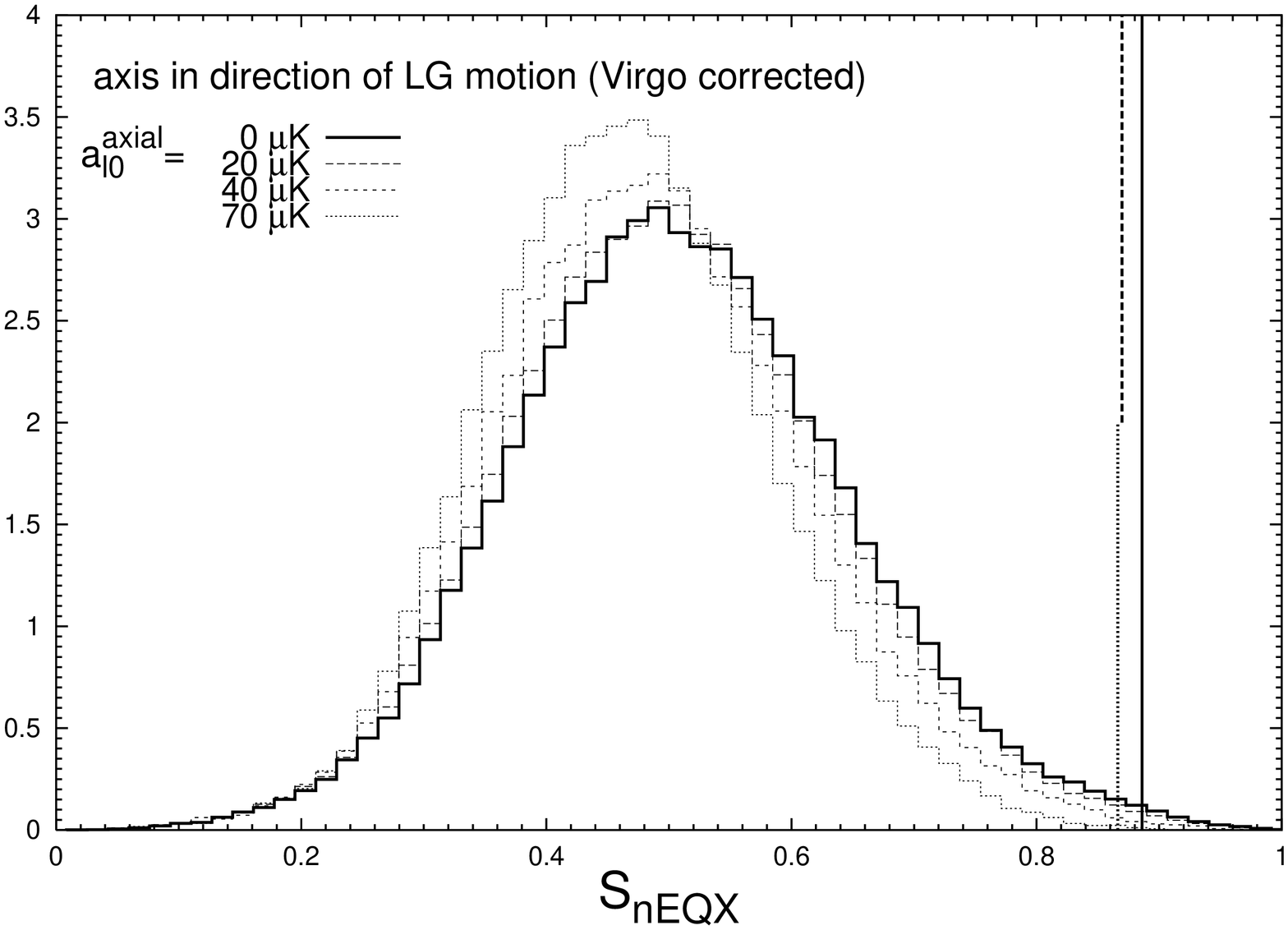}\hspace{-0.2cm}
  \includegraphics[width=1.64in,angle=-90]{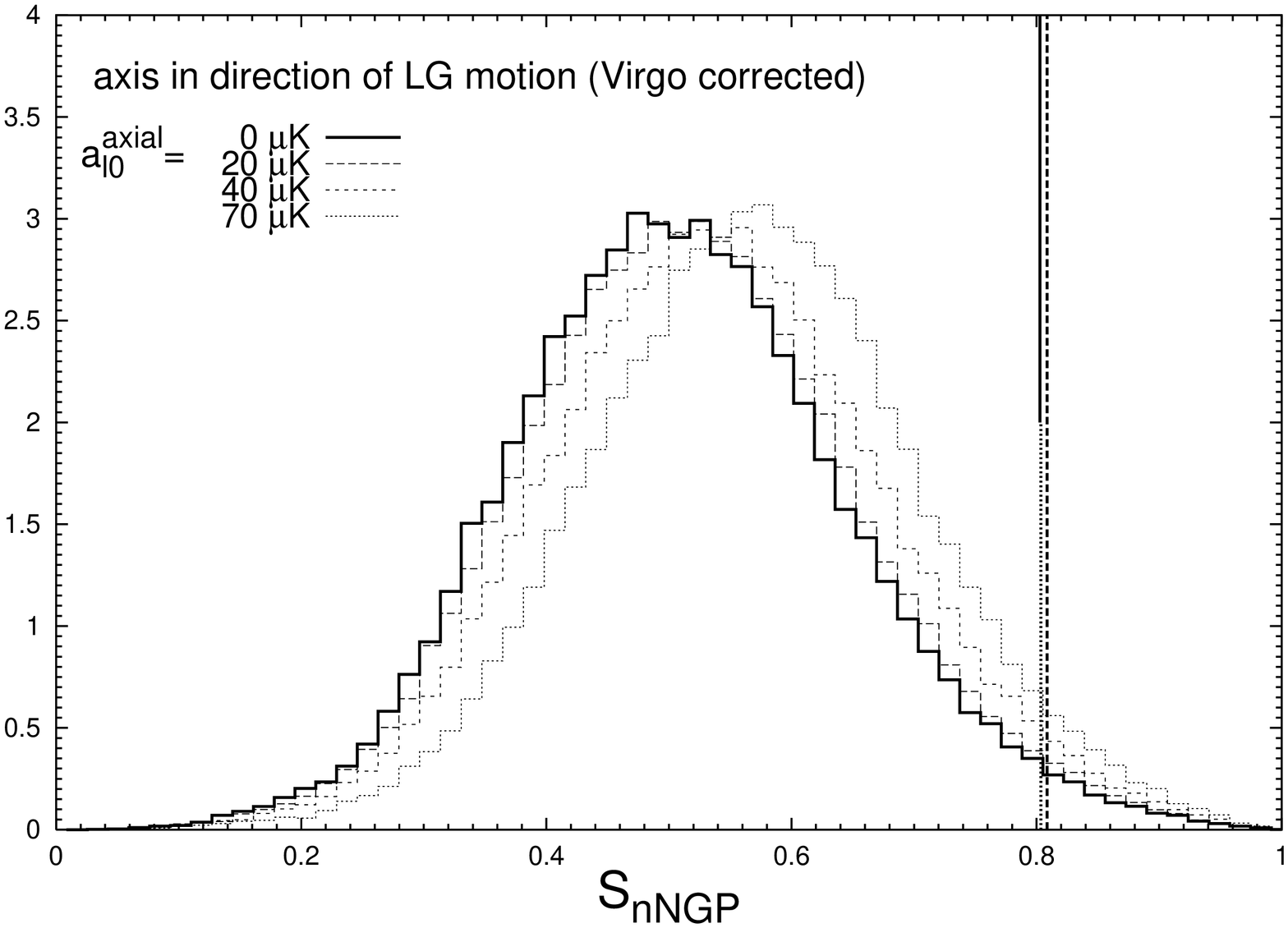}\\
  \includegraphics[width=1.64in,angle=-90]{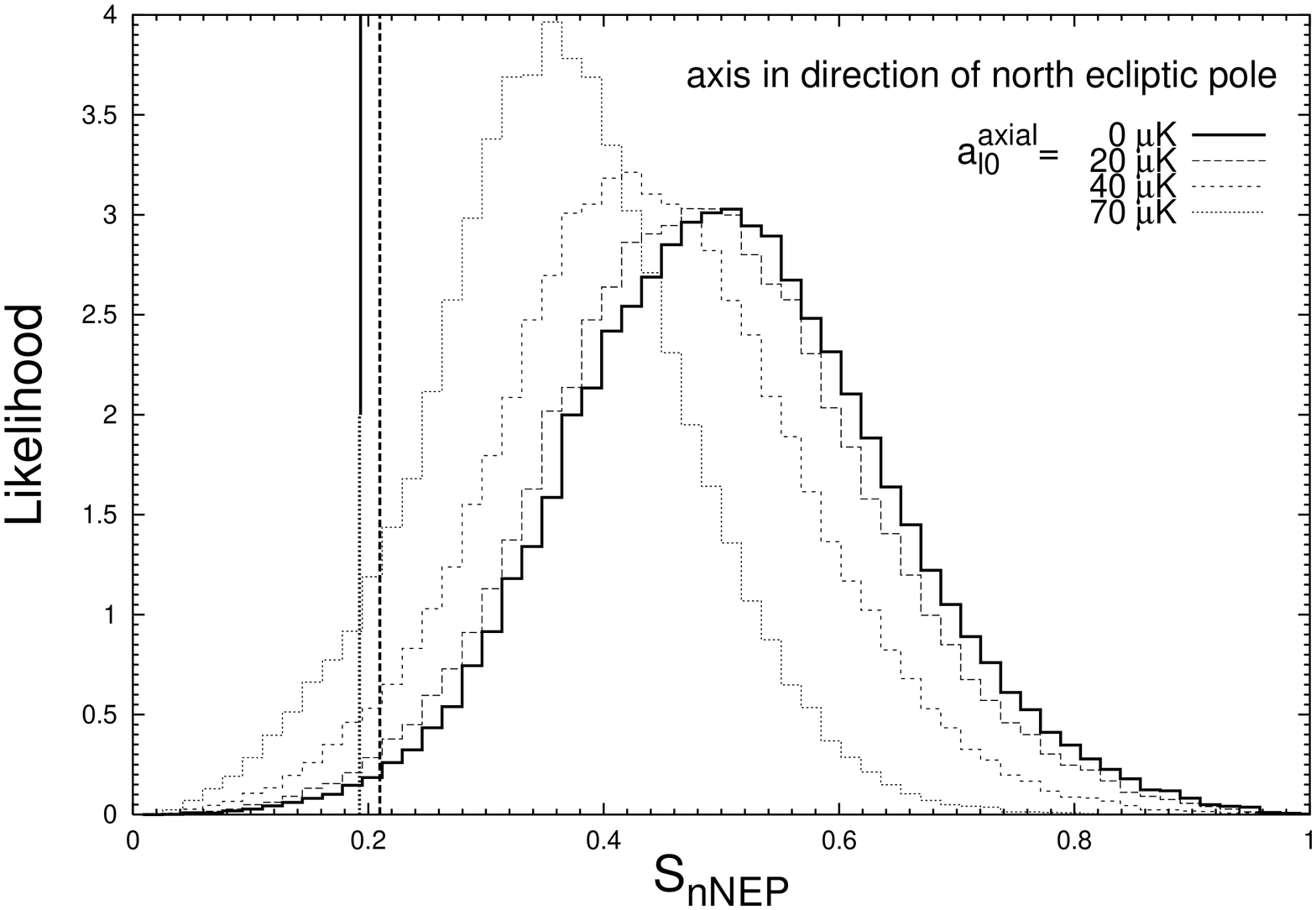}\hspace{-0.2cm}
  \includegraphics[width=1.64in,angle=-90]{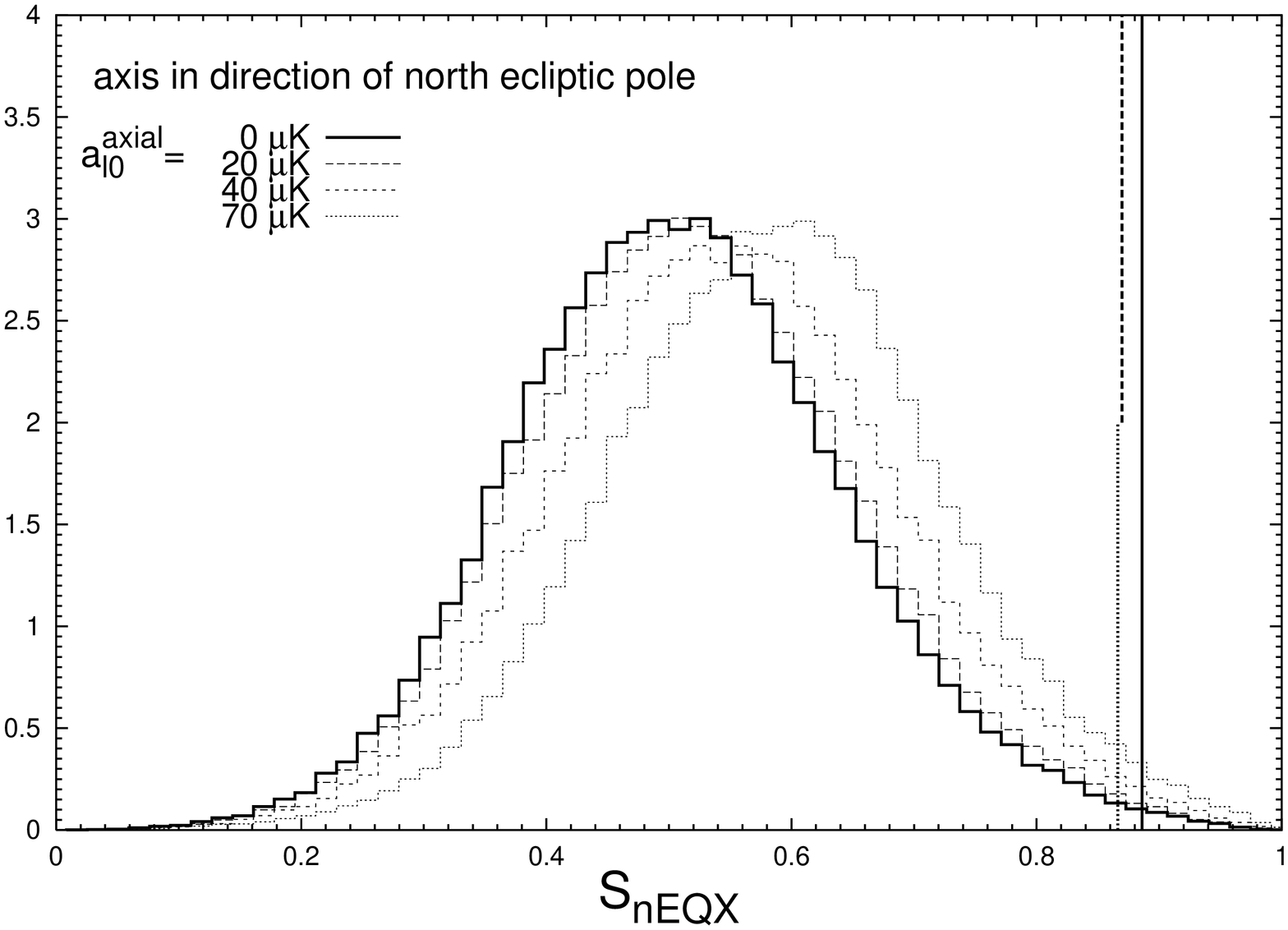}\hspace{-0.2cm}
  \includegraphics[width=1.64in,angle=-90]{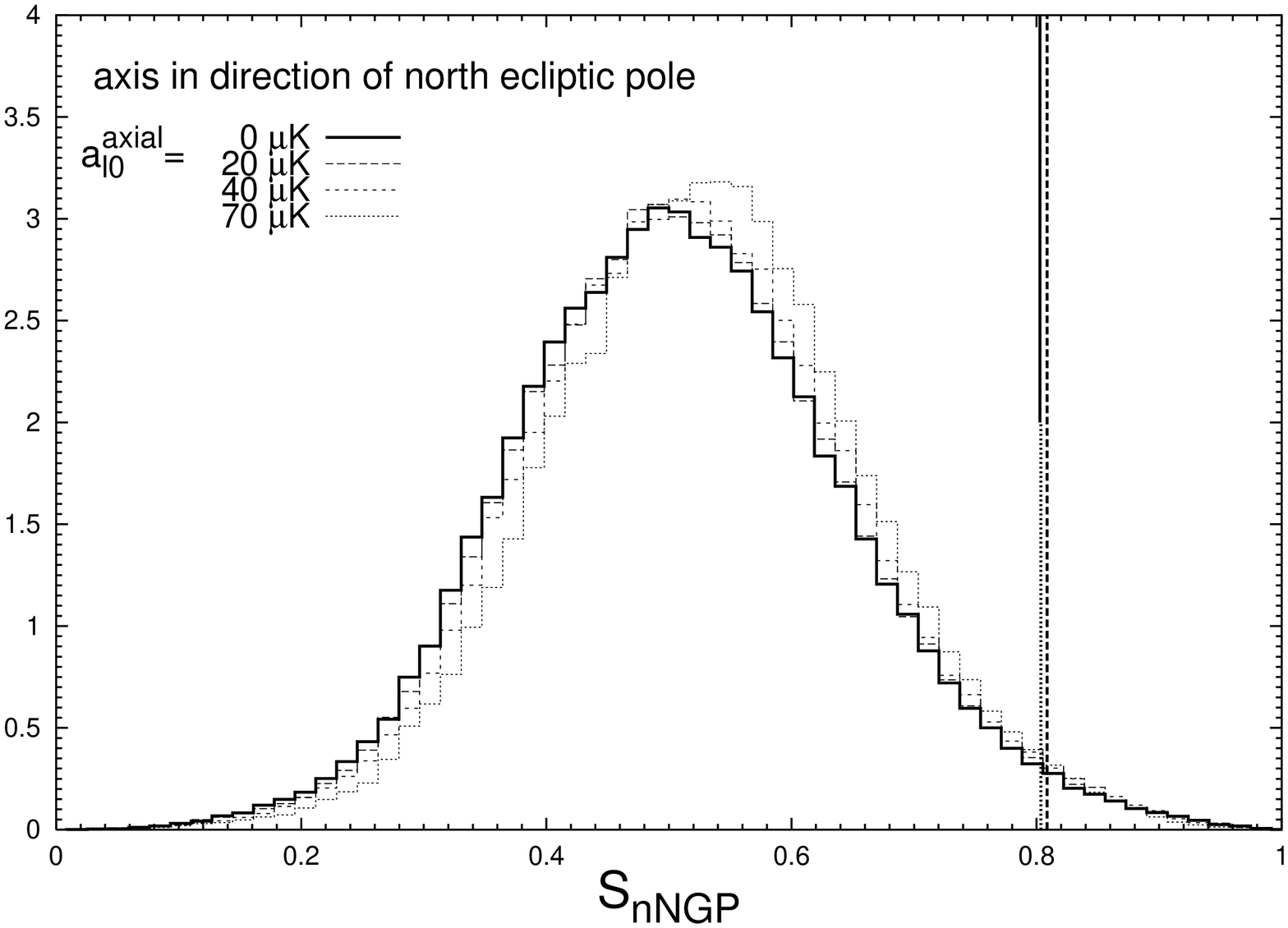}
  \caption{Alignment statistic (\ref{eq_Snx}) for quadrupole and octopole
  normals. For the three columns we pick the test direction $\hat{{\bf x}}$ to
  be NEP, EQX and NGP respectively. In rows we consider three different 
  choices of the
  preferred axis $\hat{{\bf z}}$, namely the WMAP dipole, the direction of
  motion of the LG, corrected for its Virgo infall, and the direction of the
  NEP. Shown are the likelihoods of the $S$-statistic for statistically
  isotropic gaussian skies (thick solid lines) as well as  
  different magnitudes of axial contamination of the CMB. Vertical
  lines represent the measured $S$-values from the TOH (solid line), LILC
  (dotted line) and ILC (dashed line) maps (see Table \ref{table2}). 
  Introducing a preferred axis induces correlations. For the directions
  of local motion (first and second rows) these
  correlations make the discrepancy between the measured $S$-values and model
  even bigger. At the same time, a Solar system effect is more consistent with
  data.}     
\label{fig2}
\end{figure*}

We look for alignment with three different directions $\hat{{\bf x}}$: 
the north ecliptic pole (NEP), the EQX and the north galactic pole 
(NGP). The first two are preferred directions in the Solar system
and the last defines the plane of the dominant foreground.
The observed $S$-values from the different CMB maps are given in
Table~\ref{table2}. The results of the correlation analysis are shown in 
Fig.~\ref{fig2}. By chance the CMB dipole and EQX lie very close to each
other, so an alignment test with the dipole would give results
very similar to the one with the EQX.

In the first row of Fig.~\ref{fig2} the preferred axis $\hat{\bf z}$ is 
chosen to be the measured WMAP dipole \citep{Bennett:2003a}. For all 
three tests the anomaly gets clearly worse, i.e.~the axial mechanism
drives the histograms away from the data.
Next, instead of using the motion
of the LG with respect to the CMB rest frame \citep{Kogut:1993}
as the test direction, we take the velocity of the LG
when corrected for Virgocentric motion \citep{Plionis:1998}, since this
differs more from the WMAP dipole.
The results are shown in the second row of Fig.~\ref{fig2}.
The situation for the alignment with the
EQX is again worse, but there is not much effect on
the ecliptic alignment. For the alignment
with the galactic plane, the axial contribution makes an apparent Galactic
correlation more probable, i.e.~there is a certain probability of
overestimating the galactic foreground.
For both test directions (rows one and two), the alignment with
the EQX gets worse. For example, in the direction of
the Virgo-corrected LG
motion an exclusion of $\sim 99.9\%$ C.L.\ for $a_{\ell 0}^{\rm{axial}}=50\ \mu$K can
be given for all three maps. Note that adding \emph{any}
multipole power in this test can already be excluded at the $\ge 99.4\%$ C.L.

As a complementary test we show the alignment likelihood with regard to an
orthogonal test direction, the NEP, in row three of Fig.~\ref{fig2}. 
An ecliptic extra contribution in the CMB would indeed induce an alignment of
normal vectors similar to the observed one. In particular, for
$a_{\ell 0}^{\rm{axial}}=50\ \mu$K, the probability of finding an alignment
with the NEP itself becomes roughly $5\%$, and the probability
for the EQX alignment rises to $1\%$.

To summarise, the results of recent X-ray cluster studies indicate the
existence of large-scale fluctuations $\delta \simeq \mathcal{O}(1)$
at distances of $\sim$~100~$h^{-1}$Mpc. The local RS effect
on the primordial photons from these structures is estimated to be of order
$\sim 10^{-5}$ at large angular scales. This raises 
the question of whether the local RS effect can account for the
observed anomalies in the low multipoles of the CMB.

In this letter we have assumed spherical symmetry of the
local superstructure, with an object like the Shapley Supercluster
at the centre. Under this assumption
we should observe an axisymmetric effect on the microwave sky.
The preferred axis has been taken to point in the direction of the local
velocity flow (not shown in Fig.~\ref{fig2}), the CMB dipole and the
 Virgo-corrected LG flow vector.
We have added this axisymmetric contribution to a statistically
isotropic gaussian random map and compared it by means of the
$S$-statistic with WMAP measurements.
The result is that this mechanism can be excluded at $>99\%$C.L.
Our analysis also applies to any other effect which gives an \mbox{axisymmetric} 
addition to the statistically isotropic and gaussian random sky.
However, we find that a
Solar system effect would be consistent with the data, in agreement 
with \citet{Copi:2005}. 
\begin{table}
\centering
\caption{Tests, defined in (\ref{eq_Snx}) and explained in the text, applied to
  the TOH, Lagrange ILC and ILC maps. All quadrupoles except the cut-sky 
  value have been Doppler-corrected.}  
\begin{tabular}{ccccc}
\hline 
 & Cut sky & TOH map & LILC map & ILC map \\
\hline
\hline
 $C_2$ & 129 $\mu$K$^2$ & 203 $\mu$K$^2$ & 352 $\mu$K$^2$ & 196 $\mu$K$^2$ \\
 $C_3$ & 320 $\mu$K$^2$ & 454 $\mu$K$^2$ & 571 $\mu$K$^2$ & 552 $\mu$K$^2$ \\ 
 $S_{\rm{{\bf n NEP}}}$ & - &  0.194 &  0.193 & 0.210\\
 $S_{\rm{{\bf n EQX}}}$ & - &  0.886 &  0.866 & 0.870\\
 $S_{\rm{{\bf n NGP}}}$ & - &  0.803 &  0.803 & 0.810 \\
\hline
\label{table2}
\end{tabular}
\end{table}  

The present work also indicates that the local Rees-Sciama effect
might be important for the interpretation of WMAP data and future
PLANCK data on the largest angular scales. Our work suggests
that a more detailed study of the Rees-Sciama effect could
lead to useful constraints on the local large-scale structure of the
Universe.

\hyphenation{illu-strat-ing}
\hyphenation{RS-in-du-ced}
\hyphenation{corre-lat-ions}
Supplementary figures, illustrating the RS-induced CMB correlations, are
available at http://www.physik.uni-bielefeld.de/cosmology/rs.html.
             
\section*{Acknowledgments}

We thank J.~Binney, C.~Copi, G.~Dalton, R.~Davies, D.~Huterer, S.~Sarkar,
G.~Starkman and D.~Wiltshire for discussions and comments. 
We acknowledge the use of the Legacy Archive for Microwave Background
Data Analysis (LAMBDA) provided by the NASA Office of Space Science.
The work of SR was partly done at the Rudolf Peierls Centre for
Theoretical Physics at the University of Oxford, supported by
PPARC grant PPA/G/O/2002/00479. The work of 
AR was supported by the DFG grant GRK 881.

\label{lastpage}
\end{document}